\documentclass[12pt]{article}
\usepackage{amssymb}
\usepackage{subfigure}
\usepackage{graphicx}
\usepackage{float}

\usepackage{amsmath}
\usepackage{color}

\newcommand{\bb}{\begin{equation}}
\newcommand{\ee}{\end{equation}}
\newcommand{\bega}{\begin{eqnarray}}
\newcommand{\ega}{\end{eqnarray}}
\newcommand{\begae}{\begin{eqnarray*}}
\newcommand{\egae}{\end{eqnarray*}}

\newcommand{\h}{\hspace*{4ex}}

\newcommand{\cent}{\centerline}
\newcommand{\vs}{\vspace*}

\begin{document}

\baselineskip 0.5cm

\begin{center}

{\large {\bf Experimental optical trapping of micro-particles with Frozen Waves} }

\end{center}

\vs{0.2 cm}

\cent{Rafael A. B. Suarez$^{\: 1}$, Antonio A. R. Neves$^{\: 1}$, Marcos R. R. Gesualdi$^{\: 1}$}
\cent{Leonardo A. Ambrosio$^{\: 2}$, Michel Zamboni-Rached$^{\: 3}$}

\vs{0.2 cm}

\centerline{{\em $^{\: 1}$ Universidade Federal do ABC, Av. dos Estados 5001, CEP 09210-580, Santo Andr\'e, SP, Brazil.}}
\centerline{{\em $^{\: 2}$ EESC, Universidade de São Paulo, São Carlos, SP, Brazil.}}
\centerline{{\em $^{\: 3}$ DECOM--FEEC, Universidade Estadual de Campinas, Campinas, SP, Brazil.}}

\vs{0.5 cm}

{\bf Abstract  \ --} \ In this work, we report, to the best of our knowledge, the first optical trapping experimental demonstration of micro--particles with Frozen Waves. Frozen Waves are an efficient method to model longitudinally the intensity of non--diffracting beams obtained by superposing co--propagating Bessel beams with the same frequency and order. Based on this, we investigate the optical force distribution acting on micro--particles of two types of Frozen Waves. The experimental setup of a holographic optical tweezers using spatial light modulators has been assembled and optimized. The results show that it is possible to obtain greater stability for optical trapping using Frozen waves. The significant enhancement in trapping geometry from this approach shows promising applications for optical tweezers, micro--manipulations over a broad range. \\


\vs{0.5 cm}

\h {\em\bf 1. Introduction} 

In 1986  A. Ashkin~\textit{et al}. were able to capture, three--dimensionally, dielectric particles by using a single--beam tightly focused by a high numerical aperture lens. This technique is now referred to as ``optical tweezers'' or ``optical trapping''~\cite{ashkin1986observation}. Optical tweezers have become a powerful tool for application in different fields of research, mainly in manipulation of biological system~\cite{Wang1997,Chiou2005}, colloidal systems, in nanotechnology for trapping of nano--structures~\cite{Marago2003}, in optical guiding and trapping of atoms~\cite{Kuga1997} as well as the study of mechanical properties of polymers and biopolymers. However, most applications involve the independent manipulation of multiple traps, which implies in generating complex optical traps, and thus requiring even more complex experimental systems~\cite{liesener2000multi,curtis2002dynamic}. \\

Non--diffracting waves (or diffraction--resistant waves) in optics are special optical beams that keep their intensity spatial shape during propagation. Non--diffracting beams include Bessel beams, Airy beams, and others~\cite{Durnin1987,Sivilo2007,suarez2019generation}; as well as the superposition of these waves which can produce very special structured light beams, such as the Frozen Wave (FW)~\cite{Rached2004,Rached2005,Vieira2012,Vieira2014,Vieira2015,Vieira2017,Yepes2019}. These special optical beams present great interest recently in optical tweezers, for trapping and guiding of micro-- and nano–-particles~\cite{suarez2020optimizing,ambrosio2015analytical,ambrosio2015optical}. \\

We present for the first time the experimental optical trapping of micro–particles with FWs. The transversal optical force distribution acting on micro–-particles of the two types of FWs were investigated using a holographic optical tweezers setup. Results show that is possible to obtain greater stability for optical trapping and enhances the possibility of longitudinal trapping micro-particles using FW. \\

\h {\em\bf 2. Theoretical background} 

\textbf{Frozen Waves:} About fifteen years ago, in a series of works~\cite{Rached2004,Rached2005}, an interesting theoretical method was presented, capable of providing non--diffracting beams whose longitudinal intensity patterns can be chosen at will. Such beams are obtained by suitable superposition of co--propagating Bessel beams and the resulting fields are called Frozen Wave (FW). A few years later, the experimental generation of the FW was achieved through computer generated holograms reproduced by Spatial Light Modulators (SLM)~\cite{Vieira2012,Vieira2015}. \\

The scalar Frozen Wave of zero--order is given by~\cite{Rached2004}
\begin{equation}
\Psi\left(\rho,\phi,z\right)=\sum_{n=-N}^{N}A_{n}J_{0}\left(k_{\rho_{n}}\rho\right)e^{ik_{zn}z}\,,
\label{FW_FW}
\end{equation}
where $k_{\rho n}$ and $k_{zn}$ are the transverse and longitudinal wave numbers, respectively, of the $2N+1$ Bessel beams in the superposition~\eqref{FW_FW}, and are given by
\begin{equation}
k_{\rho n}=\sqrt{2k}\sqrt{\left(k-Q\right)-\dfrac{2\pi}{L}n }, \quad k_{z n}=Q+\dfrac{2\pi}{L}n\,,
\label{krho_FW}
\end{equation}
and the coefficients $ A_{n} $ are given by
\begin{equation}
A_{n}=\frac{1}{L}\int_{0}^{L}F(z)e^{-i\frac{2\pi}{L}nz}dz.
\label{An_FW}
\end{equation}

Once the choices~\eqref{krho_FW} and~\eqref{An_FW} are made, we have that~\eqref{FW_FW} results in a beam with $|\Psi(\rho=0,z,t)| \approx |F(z)|^2$ ( where $|F(z)|^2$  is the desired longitudinal intensity pattern in the interval $0\leq z \leq L$) and with spot radius $\Delta \rho \approx 2.4 \sqrt{k^2-Q^2}$. The $Q$ parameter is a constant, where $0\leq Q \pm (2\pi/L)N \leq \omega/c$. \\

The propagation of a FW through an ABCD paraxial optical system can be described by the generalized Huygens--Fresnel diffraction integral, which in cylindrical coordinates is given by
\begin{equation}
\begin{split}
\Psi\left(\rho,\phi,z\right)& = \dfrac{ik}{2\pi B}\int_{0}^{\infty}\int_{0}^{2\pi}  \rho_{0}d\rho_{0}d\phi_{0}\Psi\left(\rho_{0},\phi_{0} \right)\\
& \times \exp\left \lbrace \dfrac{-ik}{2B} \left[A\rho^{2}_{0} - 2\rho \rho_{0}\cos\left(\phi - \phi_{0}\right) + D \rho^{2} \right]  \right\rbrace \,,
\end{split}
\label{generalized_Huygens_Fresnel_diffraction_integral_FW}
\end{equation}
where $\rho_{0}$, $\phi_{0}$ and $\rho$, $\phi$ are the radial and azimuthal angle coordinates in the input (SLM) and output planes, respectively. After some calculations we get an analytical expression for the FW of zero-order
\begin{equation}
\begin{split}
\Psi\left(\rho,\theta,z\right)=\dfrac{1}{A}&\exp\left[-i\dfrac{B}{A}\left(k-Q \right)\right]\\
& \times \sum_{n=-N}^{N}A_{n}\exp\left(-i\dfrac{2\pi B}{AL}n \right) J_{0}\left(\dfrac{k_{\rho n} \rho}{A} \right).  
\end{split}
\label{generalized_Huygens_Fresnel_diffraction_integral_FW2}
\end{equation}

These results show that FWs retain their properties when passing through an optical system, but characteristic parameters such as spot and longitudinal intensity profile can be modified. ~\cite{suarez2020propagairy,zamboni2019mathematical} \\

\h {\em\bf 3. Experiments and Results}  

\begin{figure}[H]
 \centering
  \includegraphics[width=\linewidth]{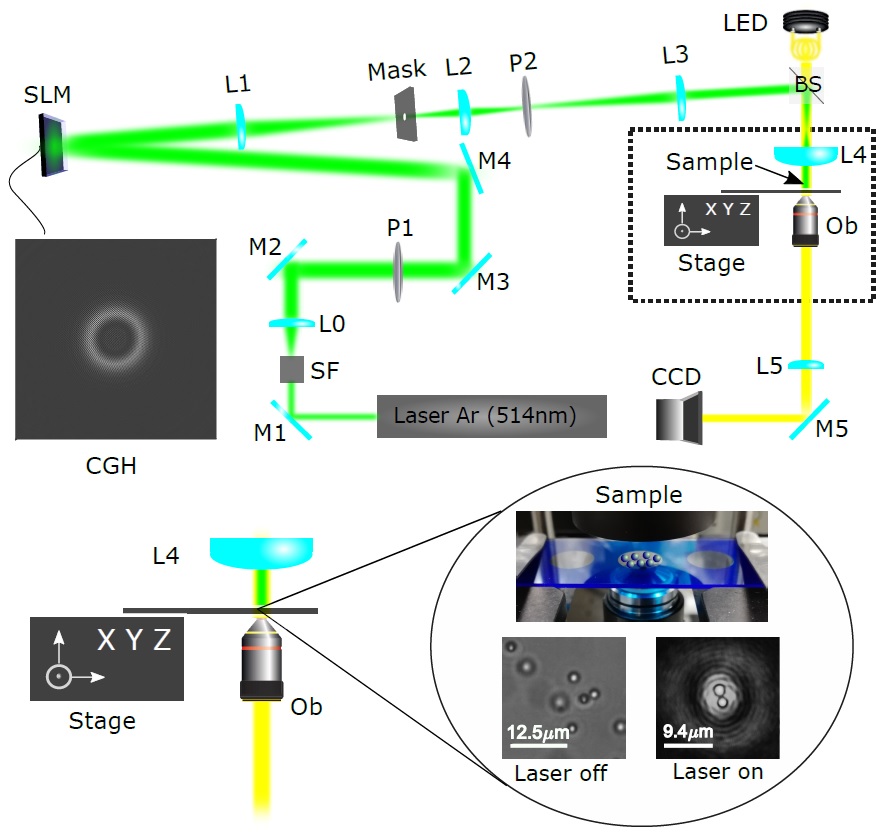}
 \caption{Experimental setup of holographic optical tweezers for optical particle trapping with non--diffracting beams using Argon laser.}
 \label{Setup_Pinza}
\end{figure}

\begin{figure}[H]
 \centering
  \includegraphics[width=\linewidth]{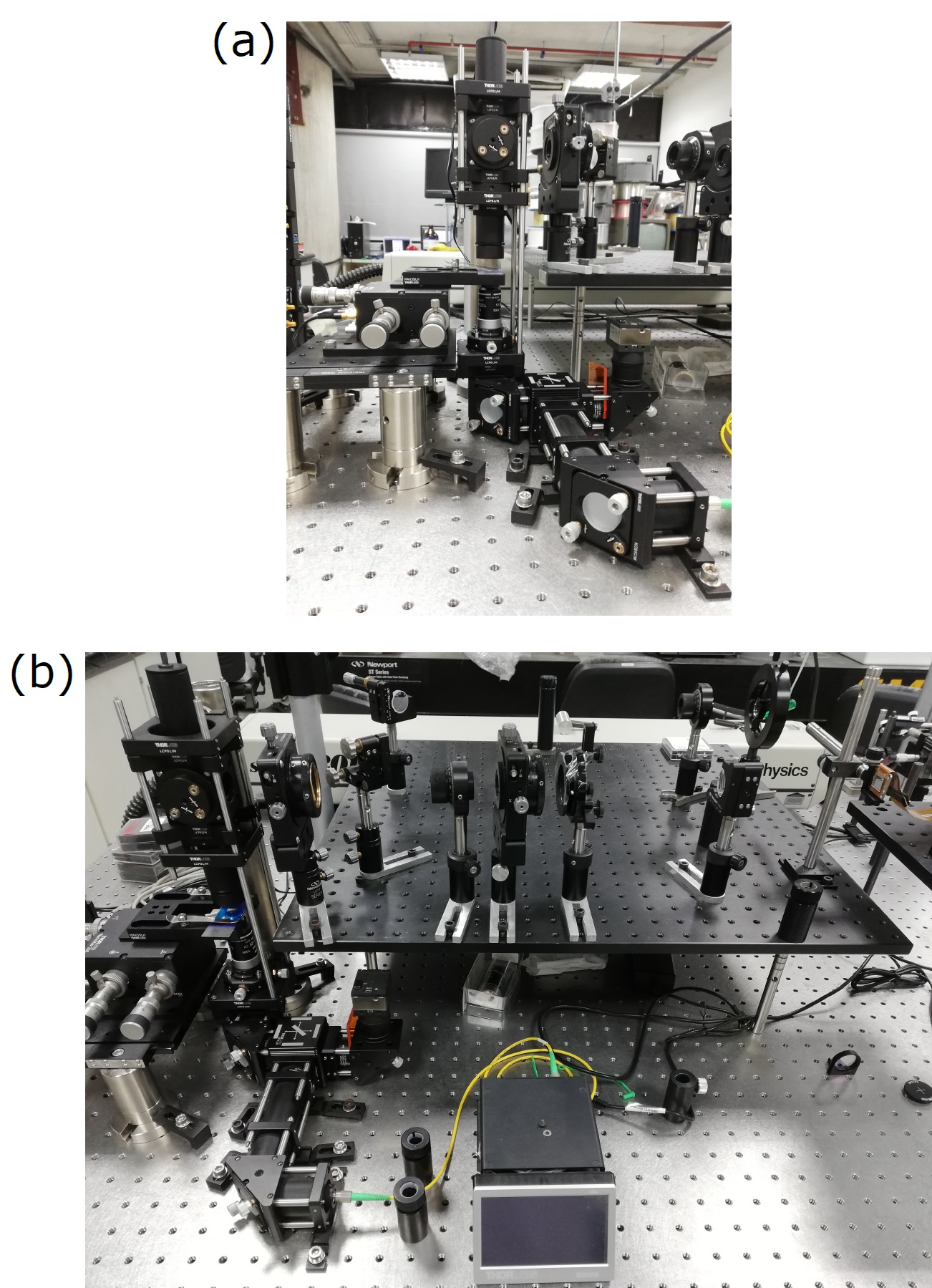}
 \caption{Holographic Optical Tweezers prototype (Photonics Lab (LabFoton), UFABC, Brazil) of the setup shown in Figure \ref{Setup_Pinza}.}
 \label{Setup_Pinza_UFABC}
\end{figure}

\textbf{Holographic optical tweezers}. In the holographic optical tweezer setup, a beam from an Argon laser with wavelength $\lambda=514.5~\text{nm} $ and output power of $300~\text{mW}$ was used. Initially, the beam passes through the spatial filter, where it is expanded and then collimated by a lens L0 with a focal length of $75~\text{mm}$ which results in a beam in diameter of about $10~\text{mm}$. The beam is directed by the M2, M3 and M4 mirrors to the spatial light modulator (LETO, Holoeye Photonics), with each pixel measuring $6.4~\mu \text{m}$ in a display matrix $1920$ $\times$ $1080$, Fig.~\ref{Setup_Pinza} and Fig.~\ref{Setup_Pinza_UFABC}. \\

The first $4f$ system consists of two lenses, L1 and L2 of focal lengths $150~\text{mm}$ and $50~\text{mm}$, respectively. On the focal plane of the lens L1 a mask was placed, references \cite{Vieira2012,Vieira2014,Vieira2015}, allowing to select the different diffraction orders in the Fourier plane of the holographically reconstructed beam. The second $4f$ system was formed with two L3 and L4 focal length lenses $150~\text{mm}$ and $25~\text{mm}$, respectively. A beam splitter BS reflects the beam vertically for the sample. The sample is placed in the focal plane of the L4 lens. To obtain the trapped particle image, the LED illumination system was used along with a $100~\times$ and $ \text{NA}=1.25$ microscope objective. \\

The optically trapped micro--particles suffer random movement due to the collisions from the surrounding fluid molecules.~\cite{Jones2015,volpe2013simulation}. Therefore, the resulting dynamics is due to the random motion and the deterministic optical forces from the intensity gradients. The process of measuring the  motion of an optically trapped particle consists of tracking its position frame by frame~\cite{helgadottir2019digital}, where each frame of the video is a two--dimensional digital image with a pixel matrix structure. 

There are several calibration methods for measuring the elastic constant, among which the most widely used are: potential analysis from the equipartition theorem, auto--correlation function analysis, and power spectrum analysis~\cite{Jones2015}. In this work, we use a method (FORMA)~\cite{garcia2018high}, which allows the reconstruction of microscopic optical force by using the Maximum Likelihood Method~\cite{kutner2005applied} and assuming that near the equilibrium position the force has a linear behavior with the particle displacement. Compared to known methods, this method has several advantages among which we have fewer samples, smaller relative errors and a high precision~\cite{garcia2018high}. \\  

\textbf{Optical Trapping with Frozen Wave}. We study experimentally, for the first time to our knowledge, the optical particle trapping process with Frozen Wave (FW). Initially, it is important to study how the propagation properties of a FW change as it passes through the two $4f$ systems, Fig.~\ref{Setup_Pinza}. As shown, from an experimental point of view it is important to choose an appropriate value for $Q$ in order to respect the SLM resolution limit~\cite{Vieira2012,Vieira2015}. We use the following values for the various FW parameters generated for optical trapping in the input plane $z=z_{\text{SLM}}$: $\lambda=514~\text{nm}$, $Q=0.999993k$, $L=2~\text{m}$, thus getting a maximum number N$_{\text{max}}=27$, which implies a spot radius $\Delta \rho=52.58~ \mu \text{m} $. When the FW optically reconstructed by SLM passes through the two $4f$ systems of Fig.~\ref{Setup_Pinza}, the propagation properties, such as spot radius and longitudinal pattern of intensity, change according to the~\eqref{generalized_Huygens_Fresnel_diffraction_integral_FW2}. For this system, the ABCD matrix is given by $M=M_{ij}$ where,  

\begin{equation}
M=\begin{pmatrix}
A & B\\ 
C & D
\end{pmatrix} = \begin{pmatrix}
\dfrac{f_{2}f_{4}}{f_{1}f_{3}} & \dfrac{f_{1}f_{3}}{f_{2}f_{4}}z\\ 
0 & \dfrac{f_{1}f_{3}}{f_{2}f_{4}}
\end{pmatrix}\,,
\label{ABCD-matrix_4f}
\end{equation}  

The~\eqref{generalized_Huygens_Fresnel_diffraction_integral_FW2} becomes
\begin{equation}
\begin{split}
\Psi &=\dfrac{f_{1}f_{3}}{f_{2}f_{4}} \exp\left[-i\left(\dfrac{f_{2}f_{4}}{f_{1}f_{3}}\right)^2 \left(k-Q \right)z\right] \sum_{n=-N}^{N}A_{n} \\ 
& \times \exp\left(-i\dfrac{2\pi n}{(f_{2}f_{4}/f_{1}f_{3})^2 L}z \right) J_{0}\left(\dfrac{k_{\rho n} \rho}{(f_{2}f_{4}/f_{1}f_{3})} \right)\,,
\end{split}
\label{FW_OT}
\end{equation}
where the new complex coefficients $A_{n}$ are given by
\begin{equation}
\begin{split}
A_{n}=\dfrac{1}{L(f_{2}f_{4}/f_{1}f_{3})^2} \int_{0}^{L(f_{2}f_{4}/f_{1}f_{3})^2}F\left[\left(\dfrac{f_{1}f_{3}}{f_{2}f_{4}} \right)^2 Z \right]& \\
\times \exp\left[-i\dfrac{2\pi n}{(f_{2}f_{4}/f_{1}f_{3})^2 L}Z \right]\text{dZ}. 
\end{split}
\label{An_OT}
\end{equation}

The~\eqref{FW_OT} and~\eqref{An_OT} revel that the spot radius is scaled by the factor  $(f_{2}f_{4}/f_{1}f_{3})$, where the longitudinal intensity pattern is scaled by $(f_{2}f_{4}/f_{1}f_{3})^2$. \\

\begin{figure}[H]
\centering
\includegraphics[scale=0.5]{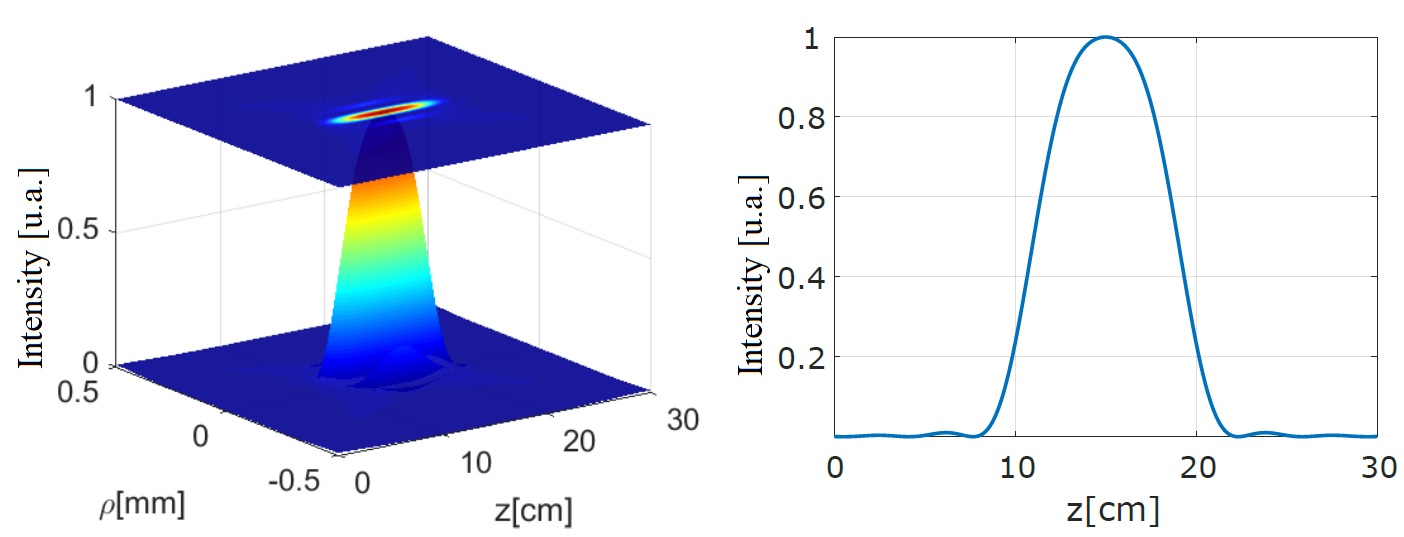}
\caption{$ (a) $ Three-dimensional profile and orthogonal projection of the theoretical normalized intensity generated in the $ z = z_{\text {SLM}} $ plane. $ (b) $ longitudinal intensity pattern (on-axis).}
\label{FW_Original_1D}
\end{figure}

\textbf{First example}. As a first example of optical trapping with FWs, let us consider that FW is generated, in the $ z=^{•}z_{\text{SLM}} $ plane, by a superposition of $(2N+1)$ zero--order Bessel beams and whose longitudinal pattern in the range $0\leq z \leq L$ is given by a step function 
\begin{equation}
F(z)= \left\{ \begin{array}{lc}
             1, &   \text{para} \quad  l_{1}\leq z \leq  l_{2} \\
             0, &   \text{elsewhere}
             \end{array}
   \right.
\label{FW_Degrau_Exp}
\end{equation}
where $l_{1}=10\text{cm}$ and $l_{2}=20\text{cm} $.
The three--dimensional profile and theoretical orthogonal projection of the normalized intensity generated on SLM output can be seen in Figure~ \ref{FW_Original_1D}. 
The corresponding longitudinal intensity pattern along the on - axis is shown in Figure~\ref{FW_Original_1D}.

\begin{figure}[H]
 \centering
  \includegraphics[width=\linewidth]{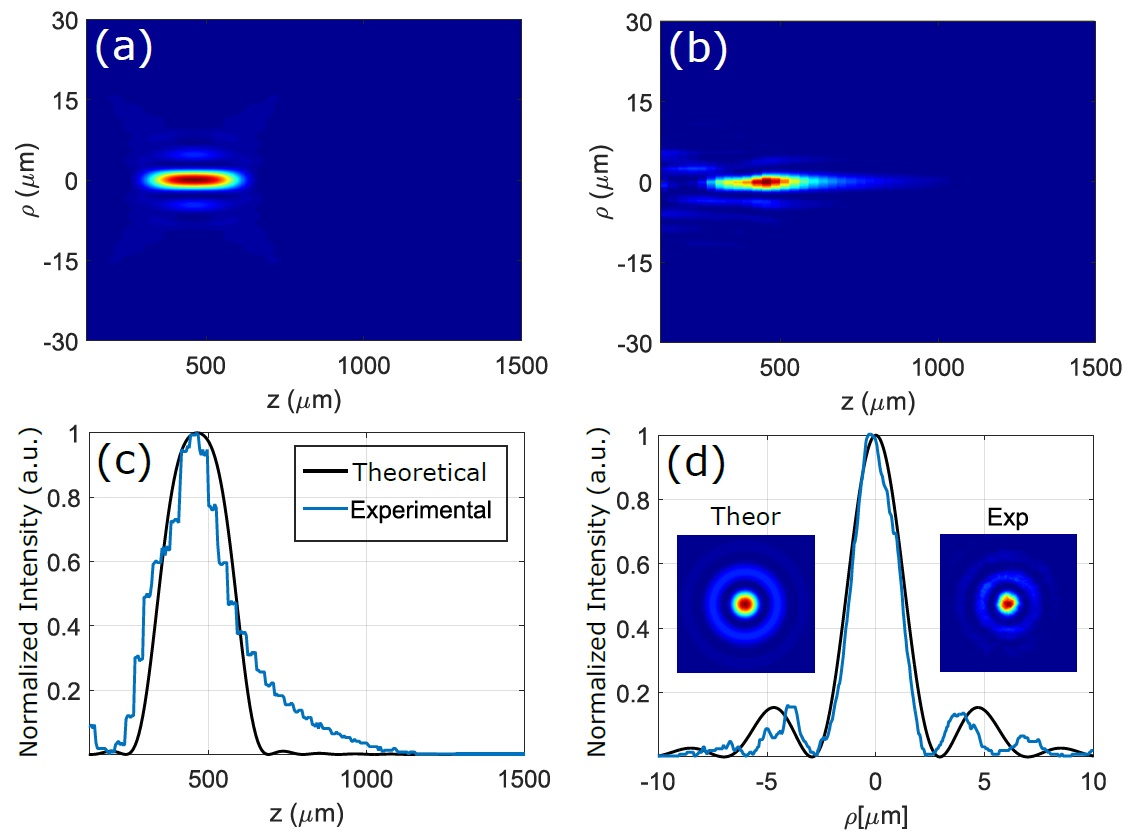}
 \caption{$(a)$ Orthogonal projection of the theoretical normalized intensity for the FW of the first example (FW1) when passing through two $4f$ systems. $(b)$ Experimental result obtained from the arrangement of Fig.~\ref{Setup_Pinza}. $(c)$ Comparison of the normalized intensity longitudinal profile between the theoretical (black line) and the experimental result (blue line). $(d)$ Normalized intensity cross--section over the plane $z_{c}$.}
 \label{FW_Transversal_Degrau}
\end{figure}

In Figure~\ref{FW_Transversal_Degrau}~$(a)$ shows orthogonal projection of the theoretical intensity as the beam passes through the $4f$ systems of the experimental arrangement of Fig.~\ref{Setup_Pinza}. The corresponding experimental result is shown in Fig.~\ref{FW_Transversal_Degrau}~$(b)$. After passing through $4f$ systems the longitudinal pattern will be set in the range~$380.6~\mu \text{m}~ \leq~z~\leq~617.3~\mu\text{m}$. In Fig.~\ref{FW_Transversal_Degrau}~$(c)$ we compare the intensity profile along the propagation axis ( $z$--axis), theoretical (black line) and experimental (blue line) axis. The normalized intensity cross--section at the center of the step $(z_{c}=462.9~\mu\text{m})$ is shown in Fig.~\ref{FW_Transversal_Degrau}~$(d)$, where the size of spot radius is $\Delta_{\rho}=2.9~\mu\text{m}$. It is important to note that the size of the $\textit{spot}$ obtained experimentally is very close to the theoretical result.

\begin{figure}[H]
 \centering
  \includegraphics[width=\linewidth]{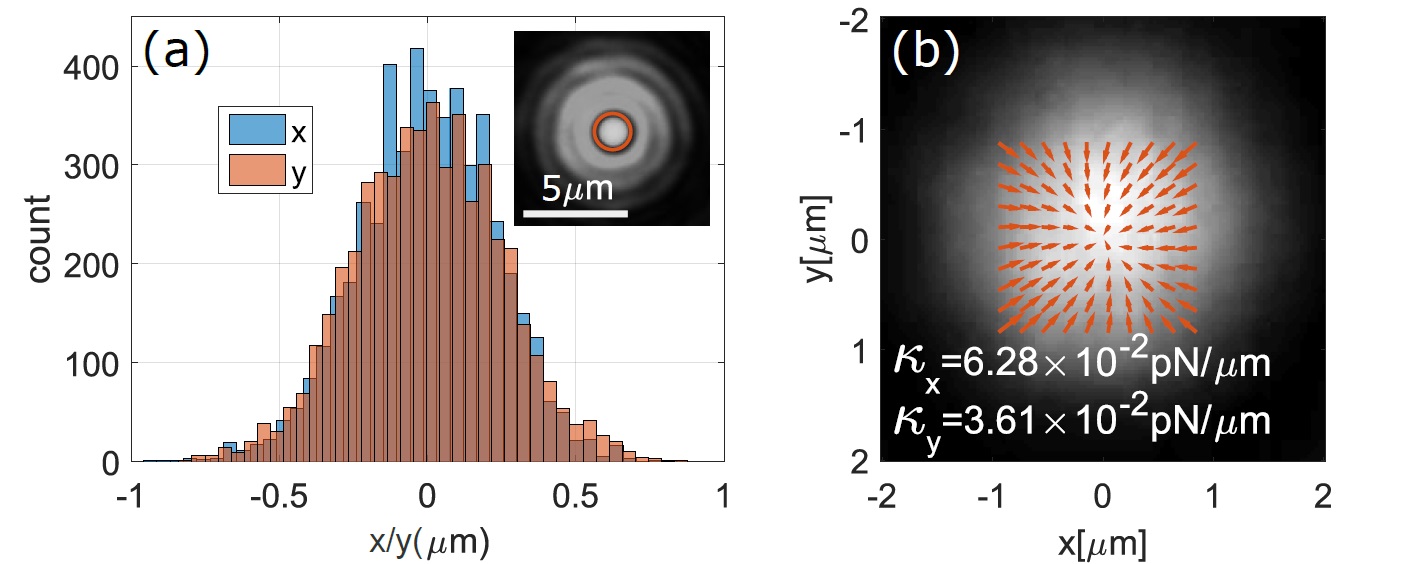}
 \caption{$(a)$ Position histograms for the $x$--axis (blue color), and --$y$ (red color) for an optically trapped particle (orange circle) at $z_{c}$. $(b)$ Force field reconstruction from the FORMA method, where $\kappa_x$ and $\kappa_y$ are elastic constant in x and y-axis, respectively. The power in the capture plane was $2.0\pm 0.5~\text{mW}$.}
 \label{Force_Distribution_Degrau}
\end{figure}

Fig.~\ref{Force_Distribution_Degrau}~$(a)$ shows the distribution of the micro--particle position (orange circle) transversely trapped by the gradient force at the center of the FW, that is, in $ (z_{c}=462.9~\mu\text{m})$. To perform the tracking, the trapped particle was recorded for $5000$ frames where the time between each frame was $17.5~\text{ms}$, and the FORMA method was used to determine the force distribution. The force field distribution can be seen in Fig.~\ref{Force_Distribution_Degrau}~$(b)$. The direction and magnitude of the arrows (orange color) correspond to the direction and magnitude of the transverse force. The intensity of the experimental cross--section of the beam in the trapping background plane is shown of the Fig.~\ref{Force_Distribution_Degrau} $(b)$. \\

\begin{figure}[H]
\centering
\includegraphics[scale=0.5]{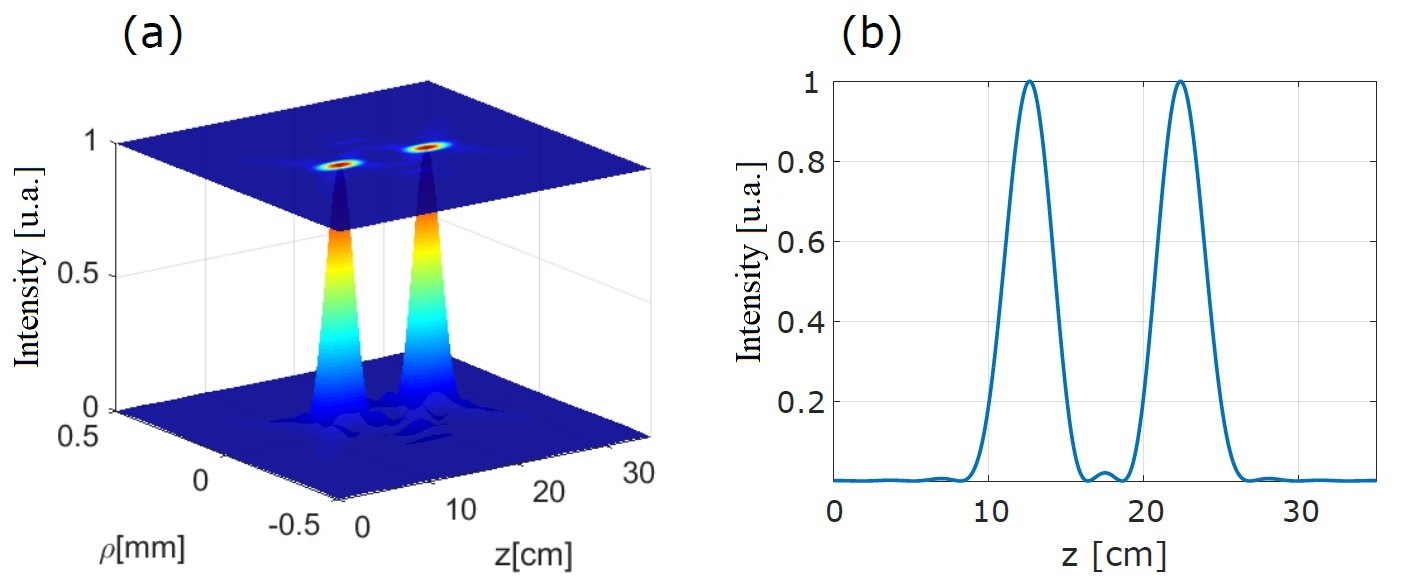}
\caption{$(a)$ Three--dimensional profile and orthogonal projection of the theoretical normalized intensity generated in the $z=z_{\text{SLM}}$ plane. $(b)$ longitudinal intensity pattern (on--axis).}
\label{FW_Original_2D}
\end{figure}

\textbf{Second Example}. As a second interesting example, let us consider a longitudinal pattern in the range $0 \leq z \leq L$ consisting of a sequence of two unitary step functions, described by
\begin{equation}
F(z)= \left\{ \begin{array}{lccc}
             1, &   \text{for} \quad  l_{1}\leq z \leq  l_{2} \\
             1, &   \text{for} \quad  l_{3}\leq z \leq  l_{4} \\
             0, &    \text{elsewhere}
             \end{array}
   \right.
\label{FW_Degrau_Exp}
\end{equation}
where $l_{1}=10~\text{cm}$, $l_{2}=15~\text{cm}$, $l_{3}=20~\text{cm}$, and $l_{4}=25~\text{cm}$. The FW generated from SLM output can be seen in Figure~\ref{FW_Original_2D}.

\begin{figure}[H]
 \centering
  \includegraphics[width=\linewidth]{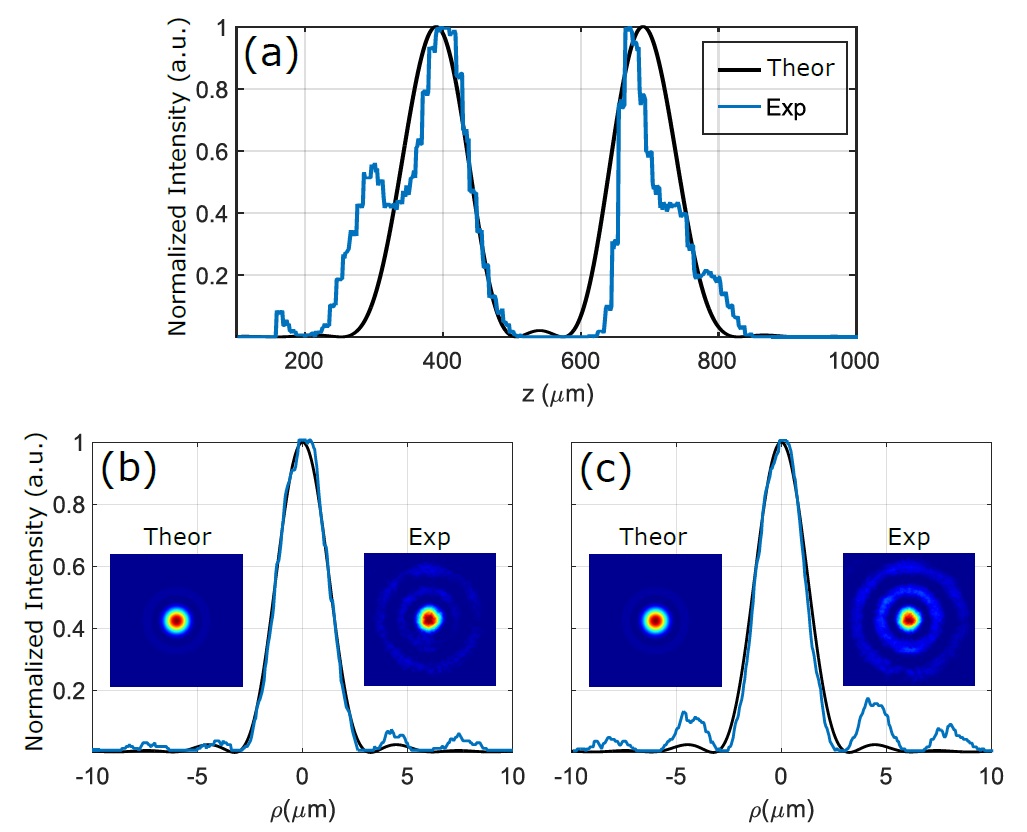}
 \caption{$(a)$ Comparison of the normalized longitudinal intensity profile between the theoretical (black line) and the experimental result (blue line) after passing through the two $4f$ systems. Cross--sectional intensity profile at the centers of the steps $(b)$ $z_{1}$ and $(c)$ $z_{2}$.}
 \label{FW_Transversal_2_Degrau}
\end{figure}

\begin{figure}[H]
 \centering
  \includegraphics[width=\linewidth]{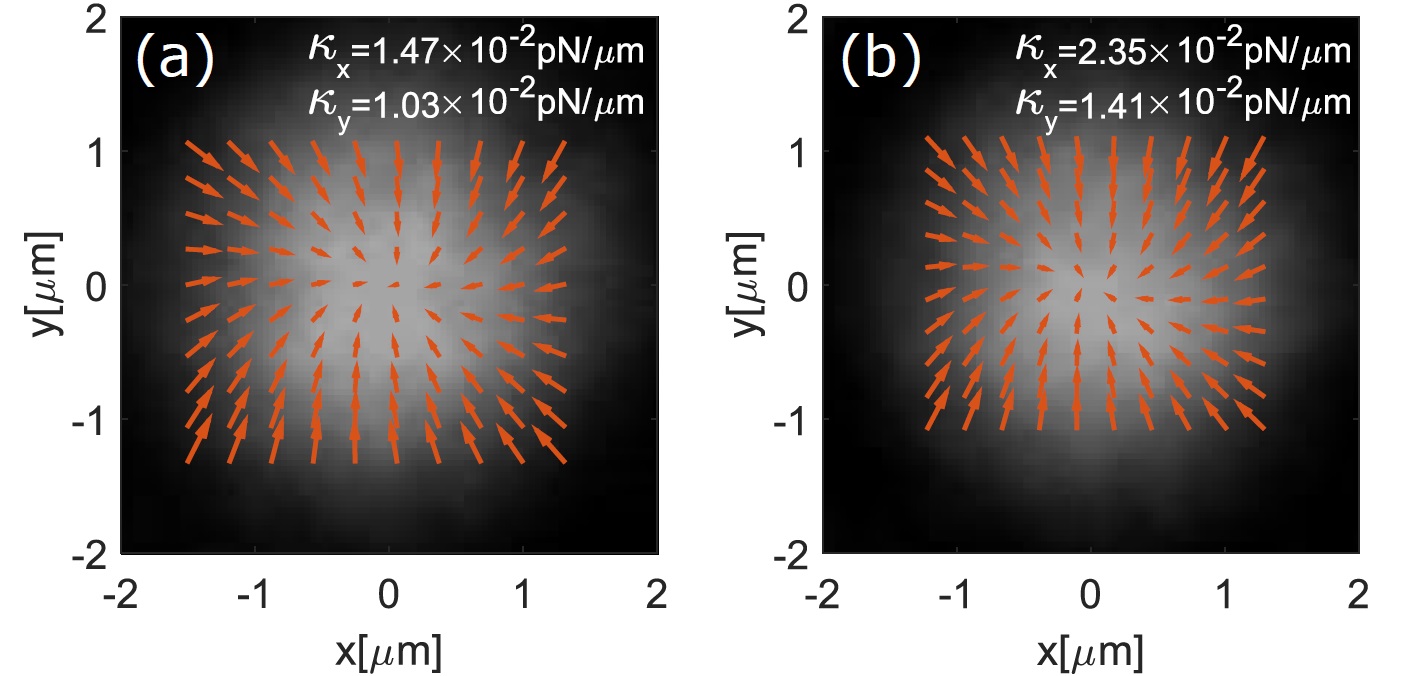}
 \caption{Force field reconstruction with second FW, consisting of a sequence of two unitary step functions, using the FORMA method. Direction and magnitude (orange arrows) corresponds to the direction and magnitude of the transverse force at the center of the steps, where $\kappa_x$ and $\kappa_y$ are elastic constant in $(a)$ $z_{1}$ and $(b)$ $z_{2}$.}
 \label{Force_Distribution_2_Degrau}
\end{figure}

After passing through the two $4f$ the FW are modified, as shown in Fig.~\ref{FW_Transversal_2_Degrau}. Figure~\ref{FW_Transversal_2_Degrau}~$(a)$ shows the comparison the theoretical intensity profile (on--axis) (black line) with the experimental result (blue line). In this case both steps are defined in the interval $308.6~\mu\text{m}~\leq~z~\leq~462.9~\mu \text{m}$ and $617.3~\mu\text{m}\leq z \leq 761.6~\mu\text{m}$, and centered on $z_{1}=389.8~\mu \text{m}$ and $z_{2}=690.3~\mu\text{m}$, respectively. In Fig.~\ref{FW_Transversal_2_Degrau} $(b)$ and $(c)$ we see the intensity cross--section on the $z_{1}$ and $z_{2}$ planes, corresponding to the centers of the steps. In both case, the $spot$ radius is very close to the theoretical result $\Delta_{\rho}=2.9~\mu\text{m}$. The distribution of the force field around the equilibrium position at the center of the steps $z_{1}$ and $z_{2}$ can be seen in Fig.~\ref{Force_Distribution_2_Degrau}~$(a)$ and $(b)$, respectively. \\

We can study how micro--particles can be trapped in different transverse planes. In this case, we use computer generated dynamic holograms to create a dynamic scene~\cite{Vieira2015} and experimentally reproduce the movement of the sample in the longitudinal direction. The latter can be done in the same way as we did to experimentally reproduce the FW, that is, keeping the static sample plane in one position and constructing a dynamic $\Psi(\rho,z_{n})$ for each value of $z$. For each value of $z_{n}$ a hologram is generated, totaling 100 holograms (frames) in the range $ [0,1000~\mu \text{m}]$. In dynamic SLM, frames are grouped to generate the dynamic beam propagation as a speed of $3$ frames/second. The data acquisition process is done keeping the sample plane static at a given position (beginning of the propagation axis), where all frames reproduced by SLM are captured.
\begin{figure}[H]
\centering
\includegraphics[width=\linewidth]{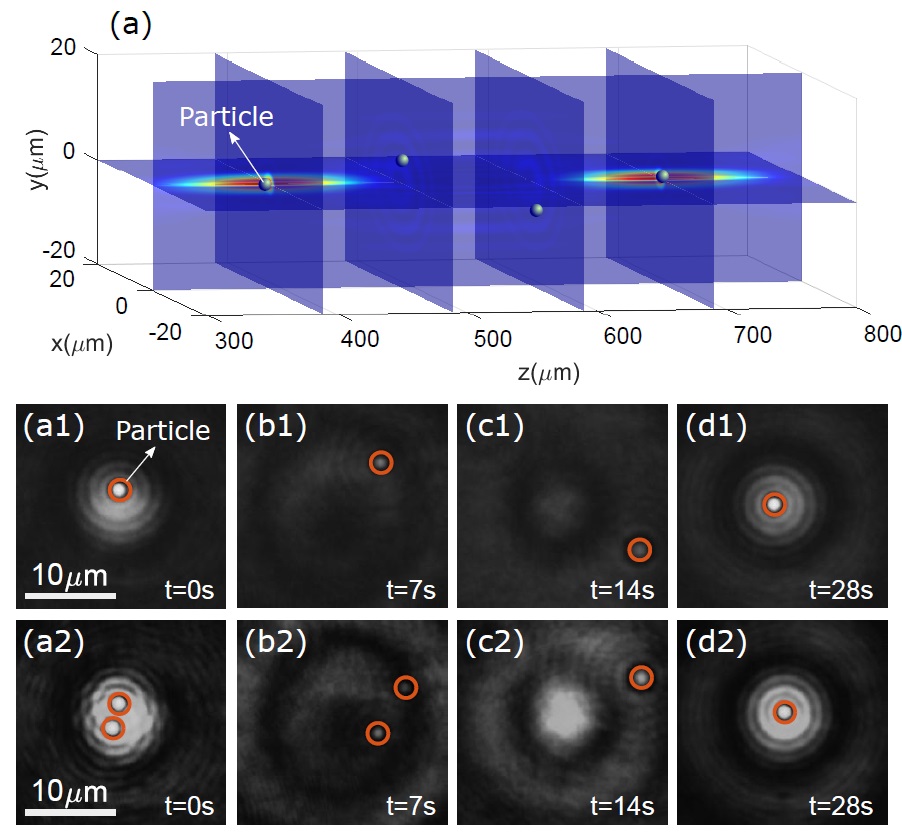}
\caption{$(a)$ Theoretical FW generated after passing through the $4f$ system for trapping and guiding the micro--particles. $(a1)$--$(d1)$ Trapping and guidance in the transverse plane of a single micro--particle (orange circle) when the FW moves at constant speed. $(a2)$--$(d2)$ Trapping and guiding in the transverse plane of two micro--particles (orange circles). The power in the capture plane was
 $1.2\pm 0.5~\text{mW}$.}
\label{FW_Trapping_1_Particle_1Degrau}
\end{figure}

In Fig.~\ref{FW_Trapping_1_Particle_1Degrau} we see the trapping and guiding of micro--particles in different transverse planes when the sample ``shifts'' along the beam propagation axis. Fig.~\ref{FW_Trapping_1_Particle_1Degrau}~$(a)$ shows the longitudinal and transverse sections of the FW, Fig.~\ref{FW_Transversal_2_Degrau} in Second Example, generated for trapping. In Figs.~ \ref{FW_Trapping_1_Particle_1Degrau}~$(a1)$--$(d1)$ we see that at the initial time $t=0$, the micro-particle is trapped in the center of the first step. However, we can observe how the micro--particle escapes from the trap when the sample exits the first step, but the micro--particle is again trapped in the center of the second step. In Fig.~\ref{FW_Trapping_1_Particle_1Degrau}~$(a2)$--$ (d2)$ we observe the same phenomenon, for the case of two trapped micro--particles on the first step, where we can see how one of them leaves the trap and does not return to the second step. \\

\h {\em\bf 4. Conclusions}  

In summary, we present for the first time the experimental optical trapping of micro--particles with Frozen Waves. The results indicate that it is possible to obtain a stable optical trap for trapping and transverse optical guidance of micro--particles using Frozen Waves. The control of longitudinal shapes by dynamic holography in an holographic optical tweezers allows to create several traps in different planes along the propagation axis. This allows transverse and longitudinal control of the trap by generating dynamic optical clamps for trapping at $3$--D. \\

\h {\em Acknowledgments} The authors acknowledge financial support from UFABC; CAPES; FAPESP (grants 16/19131-6, 15/26444-8, 17/10445-0); CNPq (grants 302070/2017-6, 307898/2018-0, 304718/2016-5). The authors also thank Tarcio A. Vieira, Mikiya Muramatsu and Erasmo Recami for their continuous collaboration.

\end{document}